\documentclass[aps,prl,twocolumn,showpacs,superscriptaddress]{revtex4-2}  

\usepackage{graphicx,xcolor}
\usepackage[utf8]{inputenc}
\usepackage{amsmath,amssymb}
\usepackage[pdfusetitle]{hyperref}
\usepackage{epstopdf}
\usepackage{slashed}

\newcommand{\lsim}{\mathrel{\mathop{\kern 0pt \rlap
  {\raise.2ex\hbox{$<$}}}
  \lower.9ex\hbox{\kern-.190em $\sim$}}}
\newcommand{\gsim}{\mathrel{\mathop{\kern 0pt \rlap
  {\raise.2ex\hbox{$>$}}}
  \lower.9ex\hbox{\kern-.190em $\sim$}}}

\definecolor{darkgreen}{HTML}{2a9000}

\newcommand{\be}{\begin{equation}}
\newcommand{\ee}{\end{equation}}
\newcommand{\ba}{\begin{array}}
\newcommand{\ea}{\end{array}}
\newcommand{\bea}{\begin{eqnarray}}
\newcommand{\eea}{\end{eqnarray}}

\newcommand{\figscale}{0.4}

\def  \bcen   {\begin{center}}
\def  \ecen   {\end{center}}
\def  \beq    {\begin{equation}}
\def  \eeq    {\end{equation}}
\def  \nn     {\nonumber }

\def\la   {\lambda}

\def\nn{\nonumber}
\def\lee { \left( }
\def\rii { \right) }
\def\lan   {\langle}
\def\ran   {\rangle}

\def\to {\rightarrow}
\def\lphp {\la^\prime_{H\Phi}}

\begin{document}
\title{
Complementary Searches of Low Mass Non-Abelian Vector Dark Matter,  \\
Dark Photon and Dark $Z^\prime$}
\author{Raymundo Ramos}
\email{raramos@gate.sinica.edu.tw}
\affiliation{Institute of Physics, Academia Sinica, Nangang, Taipei 11529, Taiwan}
\author{Van Que Tran}
\email{vqtran@nju.edu.cn}
\affiliation{School of Physics, Nanjing University, Nanjing 210093, China}
\author{Tzu-Chiang Yuan}
\email{tcyuan@phys.sinica.edu.tw}
\affiliation{Institute of Physics, Academia Sinica, Nangang, Taipei 11529, Taiwan}


\begin{abstract}
We present a novel study of the non-abelian vector dark matter candidate $W^\prime$
with a MeV$-$GeV low mass range, accompanied by a dark photon $A^\prime$ and a dark $Z^\prime$ of similar masses
in the context of a simplified gauged two-Higgs-doublet model.
The model is scrutinized by taking into account various experimental constraints including dark photon searches,
electroweak precision data, 
relic density of dark matter together with its direct and indirect searches,
mono-jet and Higgs collider physics from LHC. 
The viable parameter space of the model consistent with all experimental and theoretical constraints
is exhibited. 
While a dark $Z^\prime$ can be the dominant contribution in the relic density 
due to resonant annihilation of dark matter, a dark photon is crucial to dark matter direct detection.
We demonstrate that the parameter space can be further probed in the near future 
by sub-GeV dark matter experiments like CDEX, NEWS-G and SuperCDMS.
\end{abstract}

\maketitle

{\it Introduction.}--
Albeit mountains of experimental data, ranged from the cosmic microwave background radiation at the largest scale 
to galaxy structure formation at the local scale, are now providing strong evidences for 
the existence of dark matter (DM), its particle nature remains elusive.
The stability of massive DM is usually implemented in many particle physics models beyond the standard model (SM) 
by imposing a discrete $Z_2$ symmetry in the Lagrangian. 
The most studied DM candidate is the spin 1/2 lightest neutralino in minimal supersymmetric standard model 
with $R$-parity~\cite{Chamseddine:1982jx, Nilles:1983ge, Jungman:1995df}.

Recently a gauged two-Higgs-doublet model (G2HDM)
based on an extended electroweak gauge group ${\mathcal G} = SU(2)_L \times U(1)_Y \times SU(2)_H \times U(1)_X$
was proposed~\cite{Huang:2015wts}, in which a hidden discrete $Z_2$ symmetry ($h$-parity)~\cite{Chen:2019pnt} arises naturally 
as an accidental remnant symmetry rather than imposed {\it ad hoc} by hand. 
This discrete symmetry ensures the stability of the DM candidate in G2HDM, which can be either a complex scalar 
(in general a linear combination of various fields in the model as studied in~\cite{Chen:2019pnt})
or a heavy neutrino $\nu^H$ or an extra gauge boson $W^{\prime (p,m)} \left( W^{\prime m} = \left( W^{\prime p} \right)^* \right)$, 
all of which have odd $h$-parity. Unlike the left-right symmetric model~\cite{Mohapatra:1979ia}, 
the $W^{\prime (p,m)}$ in G2HDM do not carry electric charge.
Moreover the $h$-parity ensures there is no tree level flavor changing neutral currents 
for the SM sector in G2HDM~\cite{Huang:2015wts}.

The novel idea of G2HDM, as compared with many variants of general 2HDM~\cite{Branco:2011iw}, 
is that the two Higgs doublets $H_1$ and $H_2$ of $SU(2)_L$ are 
grouped into a fundamental representation of a new hidden $SU(2)_H$. 
Consistency checks of the model were performed in~\cite{Arhrib:2018sbz,Huang:2019obt}.
In this Letter, we will show that the non-abelian gauge boson $W^{\prime (p,m)}$ associated with  
$SU(2)_H$ can be a viable DM as well. In particular we will focus on the low mass DM scenario in the MeV$-$GeV range
and its correlations with two other neutral gauge bosons 
in $SU(2)_H \times U(1)_X$.

{\it The G2HDM Model and its Mass Spectra.}--
We begin by simplifying the original G2HDM~\cite{Huang:2015wts}
to remove the $SU(2)_H$ triplet scalar $\Delta_H$ because it is not absolutely required 
for a realistic particle spectra and the number of free parameters in the scalar potential 
can be reduced significantly.
New heavy fermions $f^{\rm H}$ are the same as before due to anomaly cancellations. 
Details of the model can be found in~\cite{Huang:2015wts,Arhrib:2018sbz,Huang:2019obt}.

The most general renormalizable Higgs potential invariant under the extended gauge group $\mathcal G$ is
\begin{align}\label{eq:V}
V = {}& - \mu^2_H   \left(H^{\alpha i}  H_{\alpha i} \right)
+  \lambda_H \left(H^{\alpha i}  H_{\alpha i} \right)^2  \nn\\
{}& + \frac{1}{2} \lambda'_H \epsilon_{\alpha \beta} \epsilon^{\gamma \delta}
\left(H^{ \alpha i}  H_{\gamma  i} \right)  \left(H^{ \beta j}  H_{\delta j} \right)  \nn \\
{}&- \mu^2_{\Phi}   \Phi_H^\dag \Phi_H  + \la_\Phi \lee \Phi_H^\dag \Phi_H  \rii^2  \\
{}& +\lambda_{H\Phi} \lee H^\dag H  \rii  \lee \Phi_H^\dag \Phi_H \rii  
 + \lambda^\prime_{H\Phi} \lee H^\dag \Phi_H  \rii  \lee \Phi_H^\dag H \rii, \nn
\end{align}
where  ($\alpha$, $\beta$, $\gamma$, $\delta$) and ($i$, $j$) refer to the $SU(2)_H$ and $SU(2)_L$ indices respectively, 
all of which run from one to two, and $H^{\alpha i} = H^*_{\alpha i}$.
$\Phi_H$ is a $SU(2)_H$ doublet.

To facilitate spontaneous symmetry breaking (SSB), 
we shift the fields based on our conventional wisdom
\begin{eqnarray}
\label{eq:scalarfields}
H_1 = 
\begin{pmatrix}
H_{11} \\ H_{12} 
\end{pmatrix}
=
\begin{pmatrix}
G^+ \\ \frac{v + h}{\sqrt 2} + i \frac{G^0}{\sqrt 2}
\end{pmatrix}
, \;
\Phi_H = 
\begin{pmatrix}
G_H^p \\ \frac{v_\Phi + \phi_2}{\sqrt 2} + i \frac{G_H^0}{\sqrt 2}
\end{pmatrix}
, \nn
\end{eqnarray}
where $v$ and $v_\Phi$ are the vacuum expectation values (VEV)s of $H_1$ and $\Phi_{H}$ fields respectively.
$H_2 = (H_{21},H_{22})^{\rm T}=( H^+, H_2^0)^{\rm T}$ is the inert doublet in G2HDM and does not have VEV.
We will work in the 't Hooft-Landau  gauge. 

The two $h$-parity even fields $h$ and $\phi_2$ mix by an angle $\theta_1$ satisfying
$\tan 2 \theta_1 =  \lambda_{H\Phi} v v_\Phi/(\lambda_\Phi v_\Phi^2 - \lambda_H v^2 )$,
gives rise to two physical states $h_1$ and $h_2$. 
 $h_1$ is identified as the 125 GeV SM-like Higgs boson and $h_2$ as a
heavier CP-even scalar boson. 
Similarly, the two $h$-parity odd complex fields $G_H^p$ and  $H_2^{0*}$ mix 
and give rise to two physical states, $\tilde G^p_H$
and $D$. $\tilde G^p_H$ is the massless Nambu-Goldstone boson absorbed 
by $W^{\prime \, p}$, while $D$ is a massive dark complex scalar.
The Goldstone bosons $G^0$, $G^\pm$ and $G^0_H$ are massless.
We note that $h_{1,2}$, $G^0$, $G^\pm$ and $G^0_H$ are even under $h$-parity, 
while $\tilde G^p_H$, $D$ and $H^\pm$ are odd~\cite{Chen:2019pnt}. 
The fact that $H^\pm$ has odd $h$-parity implies that the
$H^\pm W^\mp \gamma$ and  $H^\pm W^\mp Z$ couplings are absent in G2HDM.
  
The $W^\pm$ gauge boson of $SU(2)_L$ remains the same as in SM with its mass given by
$m_W = g v/2$. 
The $SU(2)_H$ gauge boson $W^{\prime (p,m)}$ receives mass from $\lan H_1 \ran$
and $\lan \Phi_2 \ran$, 
\beq
\label{WprimeMass}
m^2_{W'} = \frac{1}{ 4 } g^2_H\left( v^2 + v^2_\Phi \right) \; .
\eeq
As aforementioned, $W^{\prime (p,m)}$ is odd under $h$-parity, while $W^\pm$ is even. 
If $W^{\prime (p,m)}$ is the lightest $h$-parity odd particle in the model, 
it can be a stable DM candidate~\cite{WprimeDM}. 

On the other hand, the SM neutral gauge bosons $B$ and $W^3$ can mix with the 
new gauge bosons $W^{\prime 3}$ and $X$, all of which have even $h$-parity. 
Together with the Stueckelberg mass parameter $M_X$ for the abelian $U(1)_X$, SSB generates a 4 by 4 
neutral gauge boson mass matrix. 
In the basis of $\left\{ B, W^3, W^{\prime 3}, X\right\}~$\cite{Huang:2015wts, Huang:2019obt}, 
one can apply the weak rotation on upper left $2 \times 2$ block of the mass matrix,
resulting in a zero eigenvalue identified as the SM photon 
and a 3 by 3 sub-matrix which
can be diagonalized by an orthogonal rotation matrix ${\cal O}$ 
so that the physical states 
$(Z, \, Z^{\prime}, \, A^{\prime} ) = (Z^{\rm SM}, \, W^{\prime 3}, \, X )  \cdot {\mathcal O}^T$.
In this analysis, we arrange the neutral gauge boson masses as $m_{A^{\prime}} < m_{Z^{\prime}} < m_Z$
with $Z$ identified as the physical SM $Z$ boson with mass $91.1876\pm0.0021$ GeV~\cite{Zyla:2020zbs}.
Two more lighter neutral vector bosons, namely a dark photon $A^\prime$ and a dark $Z^\prime$, 
are predicted. 

The new gauge couplings $g_H$ and $g_X$ for $SU(2)_H$ and $U(1)_X$ 
are expected to be much smaller than the SM $g$ and $g'$ 
in order not to jeopardize the electroweak precision data.
The VEV $v_\Phi$ is also expected to be larger than 
$v$ since all new heavy fermion masses are proportional to it.
Furthermore, since we want the hierarchy $m_{A^{\prime}} < m_{Z^{\prime}} < m_{Z}$, 
we require $M_X < v$. The neutral gauge boson masses can then be well approximated by~\cite{InPrep} 
\begin{widetext}
\begin{equation}
\label{neutralGBMasses}
m_Z^2  \approx m^2_{Z^{\rm SM}} \equiv \frac{1}{4} \left( g^2 + g^{\prime 2} \right) v^2 \, , \; \;
m_{Z^{\prime}}^2  \approx 
	m_{W'}^2 \left(1 + \frac{4 g_X^{2}}{g_H^2}\right)
	+ M_X^{2} - m_{A^{\prime}}^2   \; \; {\rm and} \; \; 
m_{A^{\prime}}^2  \approx M_X^2 \left(1 + \frac{4 g_X^{2}}{g_H^2} + \frac{M_X^{2}}{m_{W'}^2}\right)^{-1} .
\end{equation}
\end{widetext}
Thus the masses of the DM $W^{\prime (p,m)}$, 
dark photon $A^\prime$ and dark $Z^\prime$ are entangled with each other 
and we have $m_{Z^{\prime}} \gtrsim m_{W'}$ and $m_{A^{\prime}}\lesssim M_X$. 
Since the couplings of $A^\prime$ and $Z^{\prime}$ with the SM fermions are proportional to 
the smaller couplings $g_H$ and/or $g_X$, 
the Drell-Yan type processes are suppressed and 
this can explain the null results of extra neutral gauge bosons searches at LEP\@.

It is useful to express the fundamental parameters in the scalar potential 
in terms of the physical scalar boson masses. 
Indeed one can trade six model parameters with five physical squared masses and one mixing angle,
$\{ \lambda_H, \lambda_\Phi, \lambda_{H\Phi}, \lambda'_{H\Phi}, \lambda'_H, v_\Phi \}$
$\to  \{ m^2_{h_1}, m^2_{h_2}, m^2_D, m^2_{H^\pm}, m^2_{W'}, \theta_1 \}.$
Detailed formulas for the mass spectra and these conversions will be presented elsewhere~\cite{InPrep}.
The remaining free parameters of the model are $g_H$, $g_X$, $M_X$ and $m_{f^{\rm H}}$.

{\it Theoretical Constraints.}--
(a) \underline{Vacuum Stability}: 
To make sure the scalar potential (\ref{eq:V}) is bounded from below,
we follow~\cite{Arhrib:2018sbz} and use copositivity conditions
to obtain the following constraints
$\widetilde \lambda_H (\eta) \geq 0,  \lambda_\Phi \geq 0$ and
$\widetilde \lambda_{H\Phi}(\xi) + 2 \sqrt{\widetilde \lambda_H (\eta) \lambda_\Phi}  \geq  0,$
where $\widetilde \lambda_H (\eta) \equiv \lambda_H + \eta \lambda^\prime_H$ and
${\widetilde  \lambda_{H\Phi}}(\xi) \equiv \lambda_{H\Phi} + \xi \lphp $ with $0 \leq \xi \leq 1$ and $-1 \leq \eta \leq 0$.
(b) \underline{Partial Wave Unitarity}: 
We compute all the spinless 2$\to$2 scattering
amplitudes induced by the quartic couplings in (\ref{eq:V}) 
and require their magnitudes to be less than $8 \pi$~\cite{Arhrib:2018sbz}.
(c) \underline{Electroweak Constraints}:
Following~\cite{Huang:2019obt}, we implement all the relevant constraints from electroweak precision data on the gauge sector.

{\it Dark Photon.--}
The light boson $A^{\prime}$ can be treated as a dark photon and constraints from $A^{\prime} \to \bar{\ell}\ell \, (l=e,\mu)$ should be applied.
Dark photon experiments constrain the size of the coupling
via a parameter $\varepsilon_\ell$ that appears 
as~\cite{Fabbrichesi:2020wbt}
\begin{equation}
\label{eq:Apdecay}
\Gamma\left(A^{\prime} \to \bar{\ell}\ell\right) = \frac{\alpha}{3}\varepsilon_\ell^2 m_{A^{\prime}}
\sqrt{1 - \mu_\ell^2}\left(1 + \frac{\mu_\ell^2}{2}\right),
\end{equation}
where $\mu_\ell = 2m_\ell/m_{A^{\prime}}<1$.
In the G2HDM, the parameter $\varepsilon_\ell$ at tree level is given by
\begin{equation}
\label{eq:epsilon}
\varepsilon_\ell = \frac{1}{2 s_W c_W}\sqrt{\left(v^{A^{\prime}}_\ell\right)^2
	+ \left(a^{A^{\prime}}_\ell\right)^2\left(\frac{1 - \mu_\ell^2}{1 + \mu_\ell^2/2}\right)}\;,
\end{equation}
where $v^{A^{\prime}}_\ell$ and $a^{A^{\prime}}_\ell$ are the vector and axial couplings~\cite{Huang:2019obt}.
Since $Z^{\prime}$ is also expected to be light,
the dark photon experimental limits can also be applied as above
with $A^{\prime}\to Z^{\prime}$ in~\eqref{eq:Apdecay} and $\{v_\ell^{A^{\prime}},a_\ell^{A^{\prime}}\} \to
\{v^{Z^{\prime}}_\ell,a^{Z^{\prime}}_\ell\}$ in~\eqref{eq:epsilon}.
However, since $A^{\prime}$ is lighter it is expected to be more strongly constrained.
Typically, one finds $a^{A^{\prime}}_\ell \sim 10^{-3} v^{A^{\prime}}_\ell$ and 
$a^{Z^{\prime}}_\ell \sim 10^{-2} v^{Z^{\prime}}_\ell$~\cite{InPrep}. 
Since both $v_\ell^{A^{\prime}(Z^{\prime})}$ and $a_\ell^{A^{\prime}(Z^{\prime})}$ have the same values 
for all charged leptons, $\varepsilon_\ell$ is only weakly depend on $\ell$ through $\mu_\ell$ in both cases.

Many experiments had reported stringent limits on $\varepsilon_\ell$ for low mass
$m_{A^{\prime}}$~\cite{Aaij:2019bvg,Lees:2014xha,Batley:2015lha,Banerjee:2018vgk,Riordan:1987aw,Blumlein:2011mv}. 
Current limits on $\varepsilon_\ell$ for $m_{A^{\prime}} > 1$~MeV
are displayed on the top panel of Fig.~10 in~\cite{Fabbrichesi:2020wbt}.


{\it Higgs Constraints.--}
In this analysis, we take the mass of observed Higgs boson as
$m_{h_1} = 125.10\pm0.14$~GeV~\cite{Zyla:2020zbs}.

We consider two signal strengths of $h_1 \to \gamma\gamma$ and $h_1 \to f \bar f$
from gluon fusion.
Their current experimental values are 
$\mu^{\gamma\gamma}_{\rm ggH} = 0.96 \pm 0.14$~\cite{Aad:2019mbh} and
$\mu^{\tau\tau}_{\rm ggH} = 1.05^{+0.53}_{-0.47}$~\cite{Sirunyan:2018koj}.
Besides the contributions from the SM charged particles in the 1-loop process $h_1 \to \gamma \gamma$, 
we also include all charged $f^{\rm H}$ (with $m_{f^{\rm H}}$ fixed at 3 TeV) and $H^\pm$ in G2HDM.

If $m_{h_1} > 2 \,m_{W'}$, $h_1$ can decay invisibly into a pair of $W^{\prime (p,m)}$
with invisible branching ratio ${\rm BR} ({h_1 \to {\rm inv}}) = \Gamma({h_1 \to W^{\prime p} W^{\prime m}})/\Gamma_{h_1}$.
Recently, assuming that the Higgs boson production cross section via vector boson fusion 
is comparable to the SM prediction, ATLAS sets the limit 
${\rm BR}({h_1 \to{ \rm inv}}) < 0.13$ at $95 \%$ C.L.~\cite{ATLAS:2020cjb}. 


{\it Dark Matter Constraints.}--

\noindent
\underline{Relic Density}:
The main DM annihilation channels in our model are to pairs of SM fermions
mediated by $Z$ and $Z^{\prime}$.
Other annihilation channels are also possible but are far more suppressed.
First, there is also the $A^{\prime}$ exchange diagram.
The $A^{\prime}$ couplings to SM fermions are suppressed by combinations of
new gauge couplings in $v^{A^{\prime}}_f$ and $a^{A^{\prime}}_f$.
Similarly to the case of $A^{\prime}$, the $Z^{\prime}$ couplings to the SM fermions are suppressed by its own 
$v^{Z^{\prime}}_f$ and $a^{Z^{\prime}}_f$. However, it is possible to have $\sqrt s \approx  2 m_{W^\prime}  \approx m_{Z^{\prime}}$ 
resulting in an important contribution from resonant annihilation.
Secondly, we also have the $h_1$ and $h_2$ Higgs exchange diagrams. 
Their couplings to pairs of $W^{\prime(p,m)}$ and SM fermions are
suppressed by $g_H^2$ and light fermion masses $m_q/v$ respectively.
Finally, $t$-channel diagram via exchange of $f^{\rm H}$ is suppressed by $g_H^2$ 
and the mass of the new heavy fermion in the propagator.
The annihilation mediated by the $Z$ is also suppressed by a mixing angle of ${\mathcal O}_{21}^2$
required to be small mostly by measurements on the total decay width of the $Z$ and
branching fraction for the invisible decay $Z\to W^{\prime p} W^{\prime m}$.
As mentioned above, the channel mediated by $Z^{\prime}$ is also suppressed by $g_H$ and $g_X$
inside the couplings of $v^{Z^{\prime}}_f$ and $a^{Z^{\prime}}_f$.
However, these suppressions are not as severe as in other
channels and when we include the effects from $Z^{\prime}$ resonance it is possible to
bring the relic density to its observed value of $\Omega_{\rm DM} h^2 =
0.120\pm0.001$ from Planck's measurement~\cite{Aghanim:2018eyx}.

\noindent
\underline{Direct Detection}:
Due to the small couplings between the DM candidate, $W^{\prime (p,m)}$, with
the SM-like states $h_1$ and $Z$ and the new states $h_2$, $Z^{\prime}$ and $A^{\prime}$ which
couple to the visible sector,
it is possible to have effects from DM scattering against nucleons in
detectors used in direct detection experiments.
In this case we have to consider the elastic scattering between a DM
particle and the quarks inside the nucleon.
The suppressions from $h_{1,2}$ (and $f^{\rm H}$) exchange work 
in the same way as in the processes for relic density just described.
Therefore, we are only left with the processes mediated by $Z$, $Z^{\prime}$ and $A^{\prime}$
in the $t$-channel.
Usually, for direct detection processes the momentum exchange $\vert {\bf q} \vert$ is rather small.
This will result in amplitudes suppressed by the inverse mass squared
of the mediator meaning that the lighter states, $Z^{\prime}$ and $A^{\prime}$, will be less
suppressed than the $Z$.
In the approximation that $\vert {\bf q} \vert \ll m_{Z(i)}$,
the interaction between $W^\prime$ and light quark $q$ can be written as a contact
interaction
\begin{align}
\label{eq:DDLeff}
	\mathcal{L}_{\rm CI-DD} & = \sum_q  \sum_{i=2}^3 \frac{
	g_M g_H \mathcal{O}_{2i}v^{Z(i)}_q
		}{2 m_{Z(i)}^2} \nn\\
& \;\;\;\;  \times		\left(W^{\prime p\, \mu} \partial_\nu W^{\prime m}_\mu - W^{\prime m\,
		\mu} \partial_\nu W^{\prime p}_\mu\right)\bar{q}\gamma^\nu q \; ,
\end{align}
where $Z(2)\equiv Z^{\prime}$ and $Z(3)\equiv A^{\prime}$.
It is worth noting that, as light as the mediators $Z^{\prime}$ and $A^{\prime}$ are, we can
still integrate them out thanks to the comparably small maximum momentum
transfer, $\vert {\bf q} \vert_{\max}$.
The smallness of $\vert {\bf q} \vert_{\max}$ is also due to $W'$ being light.
Consider $\vert {\bf q} \vert_{\max} = 2 \, v_{\rm DM} \, m_{W'} m_A/(m_{W'} + m_A)$ with $v_{\rm DM} = 10^{-3} c$,
$m_{W'} = 0.5$~GeV and the target mass $m_A = 131$~GeV or $40$~GeV for 
xenon or argon target respectively.
In both cases $\vert {\bf q} \vert_{\max} \sim O(1\ \text{MeV})$ while we expect $m_{A^{\prime}} \gtrsim
O(10\ \text{MeV})$ due to constraints on dark photon.  Smaller
$m_{W'}$ results in even smaller $\vert {\bf q} \vert_{\max}$.
Furthermore, for the smaller axial couplings with the quarks, in the
small momentum exchange limit, only the space components of $\gamma^\nu$
remain which are suppressed by the $W^{\prime (p,m)}$ momentum
due to the derivatives $\nabla W^{\prime (p,m)}$ as 
in~(\ref{eq:DDLeff})~\cite{Arcadi:2017kky,Escudero:2016gzx}.

From~\eqref{eq:DDLeff}, it is clear that the $A^{\prime}$ mediated process is
expected to dominate the cross section unless
$|\mathcal{O}_{23}/\mathcal{O}_{22}| < |m_{A^{\prime}}/m_{Z^{\prime}}|^2$.
The case where both mediators participate equally is expected to happen only
through fine tuning of masses and mixings.
Therefore, we expect the cross section with the nucleons to be mostly mediated
by either $A^{\prime}$ or $Z^{\prime}$.
The elastic spin-independent (SI) cross section between $W^{\prime (p,m)}$ and a
nucleon, $N$, is given by~\cite{Feng:2011vu}
\begin{align}
\label{eq:DDxsec}
\sigma^{\rm SI}_{W'N} & = 
	\sigma^{\rm SI}_{W'p}
	\frac{\sum_k \eta_k \mu_{A_k}^2 \left[Z_\text{atom} + (A_k - Z_\text{atom}) f_n/f_p\right]^2}
		{\sum_k \eta_k \mu_{A_k}^2 A_k^2} \; , \\
\label{eq:DDxsecproton}
\sigma^{\rm SI}_{W'p} & =  \frac{\mu_{p}^2 g_M^2 g_H^2 \mathcal{O}_{2i}^2}{4
\pi m_{Z(i)}^4}f_p^2 \; , \;\;\;\;\; (i = 2 \; {\rm or} \; 3)
\end{align}
where $\mu_{p} = m_{W'}m_p/(m_{W'} + m_p)$ is the reduced DM-proton mass,
$\mu_{A_k} = m_{W'}m_{A_k}/(m_{W'} + m_{A_k})$ is the reduced DM-isotope
nucleus mass, and $f_p$ and $f_n$ are effective couplings of the DM with
protons and neutrons, respectively.
$Z_\text{atom}$ is the atomic number, and the isotope dependent variables $\eta_k$ and
$A_k$ are the abundance and mass number of the $k^\text{th}$ target isotope,
respectively.
Direct detection experiments usually report the number in~\eqref{eq:DDxsec} assuming isospin conservation, 
\emph{i.e.}, $f_p = f_n$.
In that case it is straightforward to see that the ratio of the sums over
isotopes reduces to 1 and $\sigma^{\rm SI}_{W'N}  = \sigma^{\rm SI}_{W'p}$.
However, in our case the couplings between quarks, $u$ and $d$, and the gauge
bosons, $Z^{\prime}$ and $A^{\prime}$, are all different due to their distinct SM charges leading to isospin
violation (ISV), \emph{i.e.}, $f_p \neq f_n$.
Following~\cite{Feng:2011vu,Yaguna:2016bga}, we can rescale the reported experimental
limit, $\sigma_\text{limit} \to \sigma_\text{limit}\times\sigma^{\rm
SI}_{W'p}/\sigma^{\rm SI}_{W'N}$ to account for ISV effects and use it to
limit $\sigma^{\rm SI}_{W'p}$ as given by~\eqref{eq:DDxsecproton}.
This rescaling depends on the mass of DM, the atomic numbers and the
ratio $f_n/f_p$, and, hence, will be different for different points in the
parameter space.
To constraint $\sigma^{\rm SI}_{W'p}$ we use the most
recent limits set by CRESST III~\cite{Angloher:2017sxg},
DarkSide-50~\cite{Agnes:2018ves} and XENON1T~\cite{Aprile:2019xxb}.

\noindent
\underline{Indirect Detection}:
Due to DM annihilation in SM particles before freeze out happening through resonance of an
otherwise suppressed channel,
the annihilation of DM at the present time---after the shift in energy from the
early Universe to the current time---loses the resonance resulting in a very low
annihilation cross section.
We have checked that the value of the total annihilation cross section in G2HDM
at the present time is of order $10^{-32}$ cm$^3 \cdot$s$^{-1}$ or below, much lower than
the canonical limits set for various channels by Fermi-LAT data~\cite{Ackermann:2015zua, Fermi-LAT:2016uux}.

\noindent
\underline{Mono-jet}:
The event of an energetic jet with large missing transverse momentum 
has been searched by ATLAS~\cite{Aaboud:2017phn, ATLAS:2020wzf} 
and CMS~\cite{Sirunyan:2017hci} collaborations with null results.
In G2HDM, the process $ p p \to W'^p W'^m j$ can give rise to the mono-jet signal at the LHC\@. 
In the most sensitive signal region with $E_T^{\rm miss}  \in (700, 800)$~GeV 
(the signal region EM7 in~\cite{ATLAS:2020wzf}), we have checked that at 
a couple of benchmark points (BP1 and 2), the mono-jet signals 
with $p_T^{\rm miss} > 100$ GeV and precuts for jets $p_T^j > 30$ GeV and $|\eta_j| < 2.8$,
are at least two orders-of-magnitude below the current $95\%$ C.L. 
exclusion limit on the production cross section.

\begin{table}[ht]
\caption{\label{tab:prior}
Ranges and values for the prior of 8 free parameters. 
We fix $m_{f^{\rm H}} = 3$ TeV.
}
\begin{ruledtabular}
\begin{tabular}{cc} 
Parameter [units] & Range \\
\hline
$m_{h_2},\, m_D,\, m_{H^{\pm}}$ [TeV]     & [0.3\,,\,10] \\
$\log_{10}(m_{W'}/\text{GeV}),\,\log_{10}(M_X/\text{GeV})$ & [$-3$\,,\,2] \\
$\log_{10}(g_H),\,\log_{10}(g_X)$               & [$-6$\,,\,0] \\
$\theta_1$ [rad]         & [$-\pi/2$\,,\,$\pi/2$] \\
\end{tabular}
\end{ruledtabular}
\end{table}

\begin{figure}[ht]
	\includegraphics[scale=\figscale]{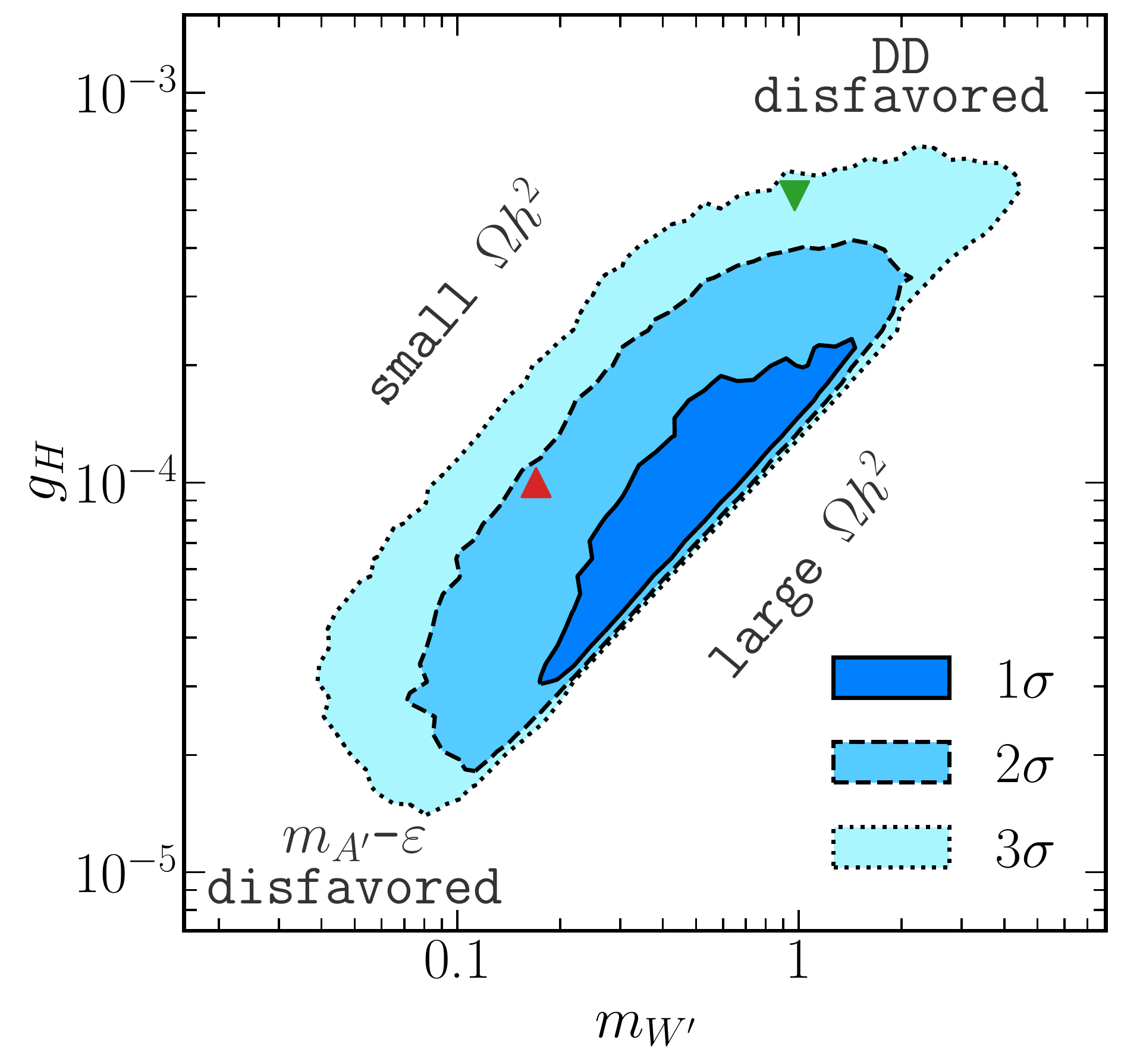}
	\caption{\label{fig:mwpgH_gHgX}
	The 1$\sigma$, 2$\sigma$ and 3$\sigma$ allowed regions 
	projected on the $(m_{W'}, g_H)$ plane.
	Benchmark points used for mono-jet analysis are also projected as indicated 
	by the green down-triangle (BP1) and red up-triangle (BP2).
	}
\end{figure}

\begin{figure*}
	\includegraphics[scale=\figscale]{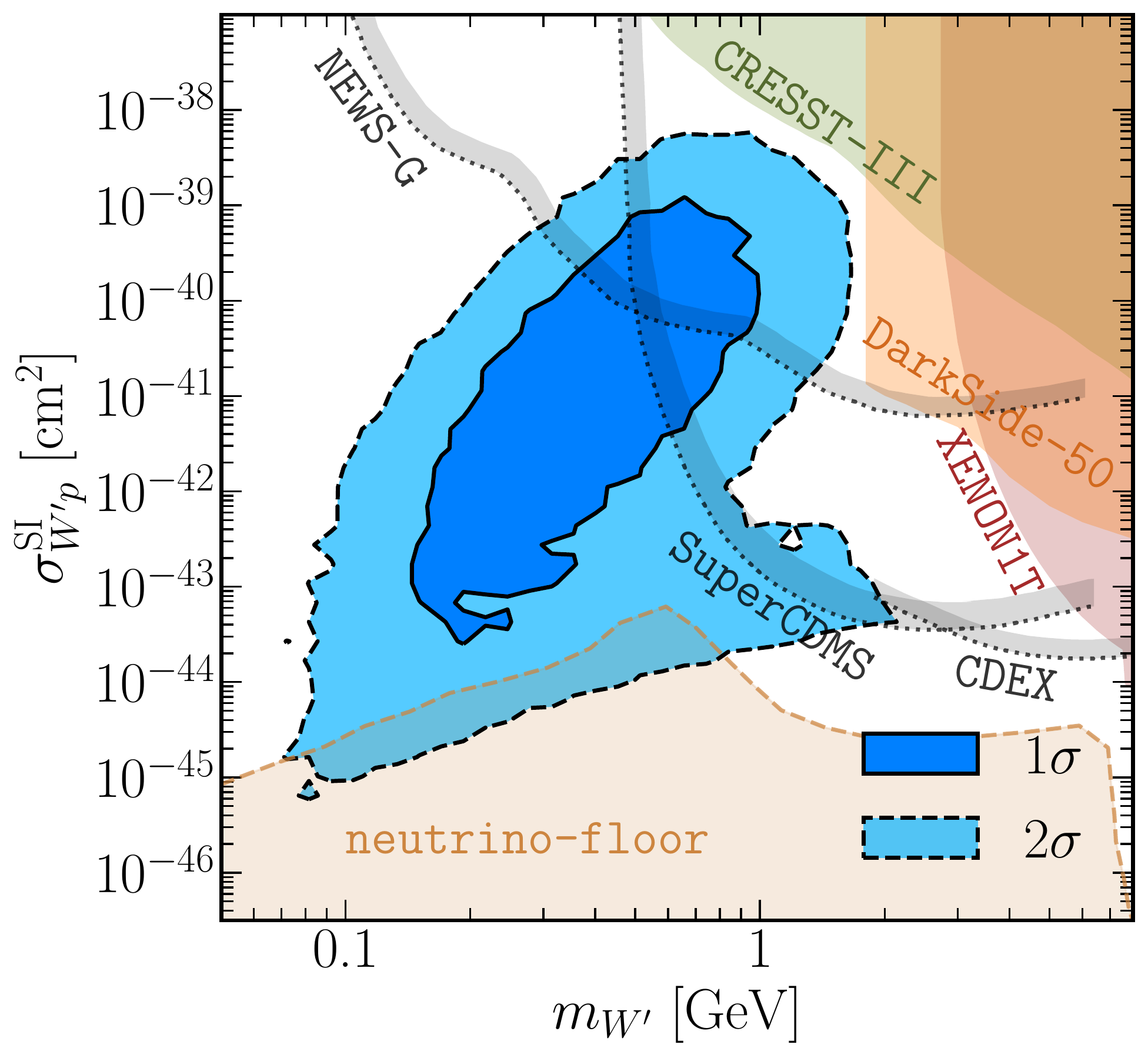}
	\includegraphics[scale=\figscale]{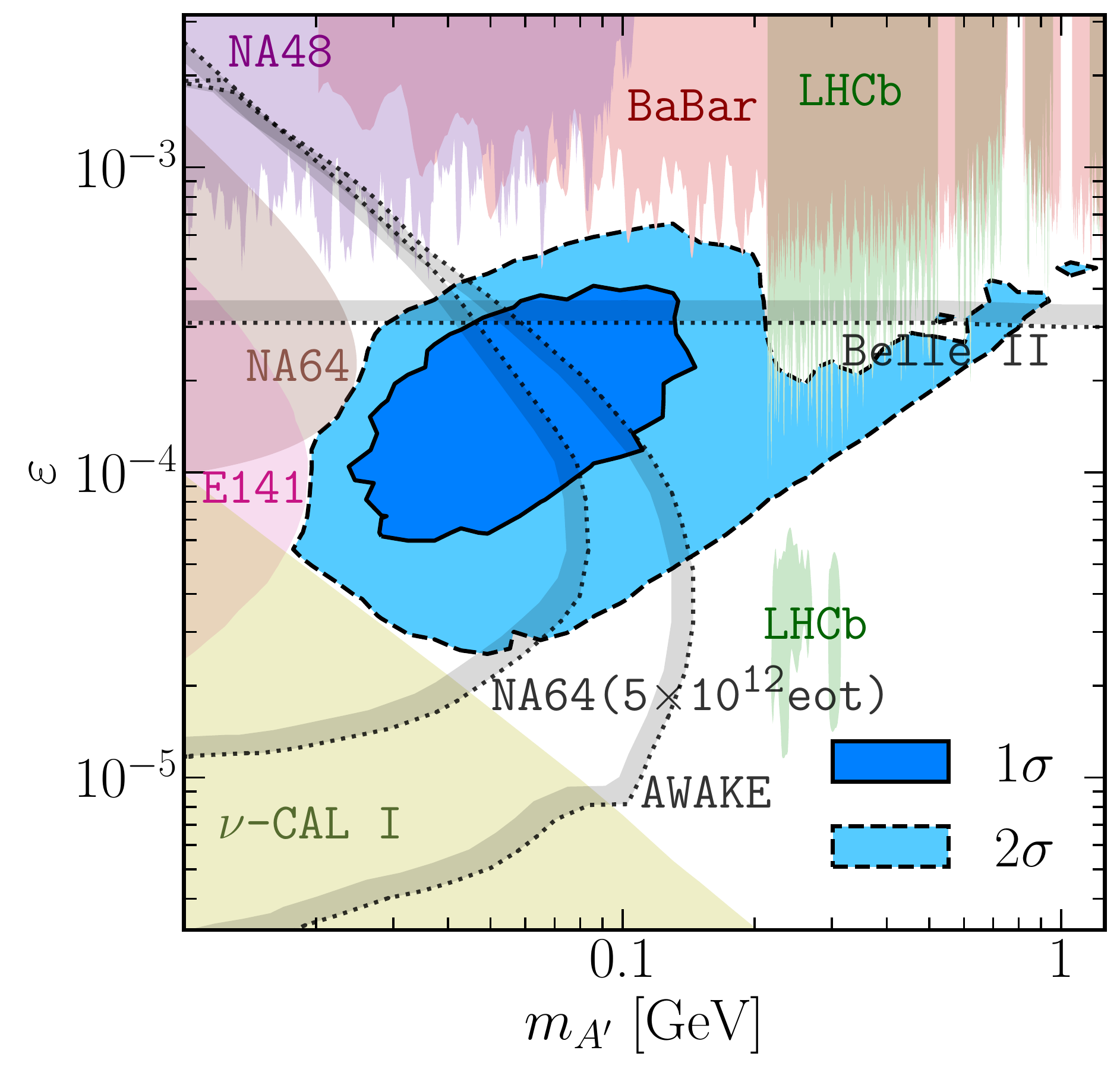}
		\caption{\label{fig:constraints}
	The 1$\sigma$ and 2$\sigma$ allowed contours
	projected on the $(m_{W'}, \sigma_{W'p}^{\rm SI})$ plane (left) 
	and the $(m_{A^{\prime}}, \varepsilon )$ plane (right).
	The experimental excluded regions used are shown as solid
	colored regions.
	Projected experimental limits are shown as dotted lines.
	}
\end{figure*}

{\it Results.}--
To sample the parameter space we use the affine invariant Markov Chain Monte
Carlo ({\tt MCMC}) ensemble sampler {\tt emcee}~\cite{ForemanMackey:2012ig} which
presents advantages such as fast calculation of parameter distributions in
multi-dimensions.
The initial priors and ranges of the free parameters used in our scan 
are tabulated in Table~\ref{tab:prior}.

The DM constraints from the relic density,
direct detection and indirect detection are calculated using
{\tt micrOMEGAs}~\cite{Belanger:2018ccd} and a set of model files generated by
{\tt FeynRules}~\cite{Alloul:2013bka}.
For the invisible decay branching ratio of the Higgs, we take advantage of
the use of {\tt CalcHEP}~\cite{Belyaev:2012qa} within {\tt micrOMEGAs} to calculate the
decay width along with the rest of the DM constraints just mentioned.

All the points outside the theoretical constraints are simply rejected.
By the same token, the dark photon constraints are used to reject any
parameter combination of $\varepsilon_{e,\mu}$-$m_{A^{\prime}}$ or
$\varepsilon_{e,\mu}$-$m_{Z^{\prime}}$ located inside the currently excluded regions.
The rest of the constraints 
are summed into a
total $\chi^2$ that also includes relic density and direct detection
SI cross section.
In the case of direct detection experiments, where a limit is reported at a
95\%~C.L.\ with null-signal assumption, we use a $\chi_\text{DD}^2$ of the form
$\chi^2_\text{DD} = 4.61\times\left(
	\sigma_\text{theory} / \sigma_\text{limit} \right)^2$, 
where the 4.61 factors allows $\chi_\text{DD}^2 = 4.61$ when we are exactly at
the 95\%~C.L.\ in 2D.
In mass ranges where more than one limit exists we take the one with the
largest $\chi_\text{DD}^2$.
Note that, due to ISV, the largest $\chi^2$ for direct detection may not
correspond to the experiment with the smallest cross section.
Since direct detection limits are reported assuming $f_p = f_n$ in~\eqref{eq:DDxsec}, 
it is possible for ISV ($f_p \neq f_n$) to produce some amount of
cancellation or enhancement of the limits depending on the atoms used in the
detector.
Computing the SI cross section in the manner described earlier in the direct detection 
allows us to account for ISV and the different atoms used in different experiments.
For the Higgs invisible decay branching fraction, 
the appropriate $\chi^2_\text{inv} = 2.71\times\left(
{\rm BR} ({h_1 \to {\rm inv}}) / 0.13 \right)^2$,
where the 2.71 factor allows for $\chi^2_\text{inv}
= 2.71$ when our result is exactly at the reported 95\%~C.L. in 1D.

Fig.~\ref{fig:mwpgH_gHgX} shows the allowed region projected on the $(m_{W'}, g_H)$ plane 
where the band-shaped region is caused by the relation 
$\Omega_{\rm DM} h^2 \propto 1/\langle \sigma v\rangle$.
Since $\sigma \propto g_H^2 m_{W'}^2/s^2$, for $s\sim 4 m_{W'}^2$ we have $g_H^2
m_{W'}^2/s^2 \sim g_H^2/16 m_{W'}^2$ resulting in $\Omega_{\rm DM} h^2 \propto
m_{W'}^2/g_H^2$.
In order to have a constant relic density, $m_{W'}$ and $g_H$ have to
maintain a linear relation as displayed in the plot.
Deviation from this band results in a relic density lying outside 
Planck's allowed range.

Note that the allowed region in Fig.~\ref{fig:mwpgH_gHgX}
are bounded in their top-right and bottom-left corners by direct detection (DD) and
the constraint on dark photon, respectively.
We know that the direct detection cross section grows with $g_H^2$ as seen in~\eqref{eq:DDxsecproton}, 
therefore, it is expected to see it setting an upper bound on $g_H$.
In the bottom-left region disfavored by dark photon searches, this is mostly
due to the constraint from $\nu$-CAL I which limits $\varepsilon$ from below as 
will be explained in the next figure. 

In the left panel of Fig.~\ref{fig:constraints} we show the allowed region projected on
the ($m_{W'}$, $\sigma_{W'p}^{\rm SI}$) plane.
The dark (light) blue shaded zone represents the $1\sigma$ ($2\sigma$)
allowed region.
The current DM direct detection measurements from
CRESST III (green)~\cite{Angloher:2017sxg},
DarkSide-50 (orange)~\cite{Agnes:2018ves} 
and XENON1T (brown)~\cite{Aprile:2019xxb}
constrain the DM mass to remain below $\sim 2$~GeV.
A small part of the $2 \sigma$ allowed region lies below the neutrino floor (light orange),
where the coherent neutrino-nucleus scattering would dominate over any DM signal.
Additionally, we show that experiments in the near future such as
NEWS-G~\cite{Battaglieri:2017aum}, SuperCDMS~\cite{Agnese:2016cpb} and CDEX~\cite{Ma:2017nhc}
can further probe our allowed parameter space, in particular 
$m_{W'}$ can reach $\gsim 0.3$~GeV with NEWS-G and $\sigma_{W'p}^{\rm SI}$ can push down to about
$10^{-44}$~cm$^2$ with SuperCDMS and CDEX\@.

The right panel in Fig.~\ref{fig:constraints} shows the allowed region projected on
the ($m_{A^{\prime}}$, $\varepsilon$) plane.
Various experimental limits from dark photon searches are displayed in color shaded zones including
LHCb (green)~\cite{Aaij:2019bvg}, BaBar (pink)~\cite{Lees:2014xha},
NA48 (purple)~\cite{Batley:2015lha}, NA64 (light brown)~\cite{Banerjee:2018vgk},
E141 (magenta)~\cite{Riordan:1987aw} and $\nu$-CAL I (olive green)~\cite{Blumlein:2011mv}.
The dilepton searches at the LHCb, BaBar and NA48 put upper limits of
$\varepsilon \lesssim 10^{-3}$ for $m_{A^{\prime}} \gsim 0.03$ GeV, especially LHCb
which sets a strong limit on $\varepsilon$ at $0.2$~GeV $< m_{A^{\prime}} < 0.5$~GeV
causing a concave region 
in the $2 \sigma$ allowed region at this mass range.
We note that this concave region due to LHCb corresponds to
the right-tilted concave region at $(m_W', \sigma_{W'p}^{\rm SI}) \sim (1 \rm\,{GeV},
10^{-42}\, \rm{cm^2})$ in the left panel of the same figure. 
Experimental probes of dark photon, dark $Z^\prime$ 
and dark matter are thus correlated.
The LHCb long lived dark photon search constraints~\cite{Aaij:2019bvg} are also shown by the two isolated green shaded islands.

On the other hand, the beam dump experiments NA64,
E141 and $\nu$-CAL I close the available space for smaller $\varepsilon$ and lighter
$m_{A^{\prime}}$ setting lower bounds of $m_{A^{\prime}} > 0.02$~GeV
and $\varepsilon \gsim 2\times 10^{-5}$.
The lower limit on $\varepsilon$ for $m_{A^{\prime}} > 0.05$ GeV
is due to the DM relic density measured by the Planck's experiment.

Interestingly, our final allowed region is located in
the gap between the beam-dump and the collider based experiments,
an area of special interest for future dark photon searches.
For instance, Belle-II~\cite{Kou:2018nap} with a luminosity of
$50\ {\rm ab}^{-1}$ can probe $\varepsilon$ down to $2 \times 10^{-4}$, 
the next upgrade of NA64~\cite{NA64:2018} can cover
$10^{-5} \lsim \varepsilon \lsim 10^{-3}$ and $m_{A^{\prime}} \lsim0.08$~GeV
by reaching $\sim 5 \times 10^{12}$ electrons-on-target (abbreviated by eot in the figure)
and Advanced WAKEfield Experiment (AWAKE) run 2~\cite{Caldwell:2018atq}
can reach $m_{A^{\prime}}$ up to $0.15$~GeV with $10^{16}$
electrons-on-target with an energy of 50~GeV.
These limits are shown explicitly in the right panel of
Fig.~\ref{fig:constraints}.
In the future, with access to high energy electron-proton colliders, AWAKE may
reach 1~TeV for the electrons, extending $m_{A^{\prime}}$ up to
0.6~GeV~\cite{Caldwell:2018atq}.

{\it Conclusions.}--
To summarize, we found that the simplified G2HDM developed in this work
provides a viable vector DM candidate $W^{\prime}$ with mass down to
$O (10^{-2})$~GeV\@.
All the predictions in the model are in good agreement with
current observations.
Importantly, both new vector states, $Z'$ and $A'$, play key roles for DM observables. 
They are the portal connecting the visible SM and hidden sectors 
via superweak interactions with couplings $g_H$ and $g_X$ of size $O(10^{-5}-10^{-3})$.   
While the dark $Z^\prime$ can be the dominant resonant contribution for DM relic density, 
the dark photon is crucial for DM direct detection.
Besides the possibility of detecting a low mass $W^{\prime}$ in DM direct
detection experiments, the dark photon $A^\prime$ is predicted to be well
positioned for future observations that may reach $m_{A'} \sim 0.1$~GeV\@.
This work demonstrates that the G2HDM is a successful and competitive dark matter
model with diverse exploration possibilities. 

We conclude that experimental searches for low mass non-abelian $W^{\prime}$ DM, 
dark photon and dark $Z^\prime$ in the sub-GeV range are complimentary with each other. 
Although our analysis is carried
out within a specific model, we expect our results may be generic and are of general interests.

\medskip

This work is supported in part by the Ministry of Science and Technology of 
Taiwan under Grant Nos. 108-2112-M-001-018 (TCY) and 108-2811-M-001-550 (RR) and 
by National Natural Science Foundation of China under Grant Nos. 11775109 and U1738134 (VQT).

\end{document}